\documentclass[12pt,aps,showpacs]{revtex4}
\usepackage{graphicx}
\begin{document}
\title{Minimization of nonresonant effects in a scalable
Ising spin quantum computer}
\author{G.P. Berman$^1$, D.I. Kamenev$^1$, and V.I.
Tsifrinovich$^{2}$}
 \affiliation{$^1$Theoretical Division and
Center for Nonlinear Studies, Los Alamos National Laboratory, Los
Alamos, New Mexico 87545}
\affiliation{$^2$ IDS Department, Polytechnic University, Six
Metrotech Center, Brooklyn, New York 11201}
\vspace{3mm}
\begin{abstract}

The errors caused by the transitions with large frequency offsets
(nonresonant transitions) are calculated analytically for a
scalable solid-state quantum computer based on a one-dimensional
spin chain with Ising interactions between neighboring
spins. Selective excitations of the spins are enabled by 
a uniform gradient of the external magnetic field. We calculate the
probabilities of all unwanted nonresonant transitions associated with the
flip of each spin with nonresonant frequency and with flips of
two spins: one with resonant and one with nonresonant
frequencies. It is shown that these errors oscillate with changing
the gradient of the external magnetic field. Choosing the optimal
values of this gradient allows us to decrease these errors by 50\%.
\end{abstract}
\maketitle

\section{Introduction}

Even with no interaction with the environment,
errors in a quantum computer driven by radio-frequency
({\it rf}) pulses may be generated by the nonresonant action of these
pulses on different spins. (See, for example, \cite{Lopez,Method}.)
These errors depend on the value of the detuning from the resonant
(useful) transitions. Unwanted transitions with small
detunings produce the largest errors. Such transitions, called 
near-resonant transitions, are associated with
interactions between the spins in the chain. 
These interactions are required in order to implement the
conditional quantum logic. But the same
interactions can cause unwanted transitions characterized by 
small detunings from resonance and relatively large probabilities.  
If one wants to flip the $k$th spin in a particular state and
does not want to flip the same spin in the other state with
different orientations of the neighboring $(k+1)$th and $(k-1)$th
spins, then first of all one has to suppress these near-resonant
transitions. For example, if the resonant
transition is
\begin{equation}
\label{transition}
|\dots0_{k-1}0_k1_{k+1}\dots\rangle\rightarrow
|\dots0_{k-1}1_k1_{k+1}\dots\rangle,
\end{equation}
then the (unwanted) near-resonant transitions to be suppressed are
\begin{equation}
\label{transition1}
|\dots0_{k-1}0_k0_{k+1}\dots\rangle\rightarrow
|\dots0_{k-1}1_k0_{k+1}\dots\rangle,~~~
|\dots1_{k-1}0_k1_{k+1}\dots\rangle\rightarrow
|\dots1_{k-1}1_k1_{k+1}\dots\rangle.
\end{equation}
A general approach for the complete suppression of 
near-resonant transitions for an arbitrary superposition of
quantum sates in a scalable Ising spin quantum computer was
developed in \cite{Method}.

Since a wavelength of a {\it rf} pulse is much larger than a
distance between the spins, the pulse affects all spins in the
chain. Hence, in order to allow a selective excitation of the
$k$th spin, one must make the Larmor frequency of this 
particular spin different from the Larmor frequencies of all other
spins, and choose the frequency of the {\it rf}  pulse to be equal
to the transition frequency of this $k$th spin. In Kane's
\cite{Kane} semiconductor quantum computer proposal this can be done by
applying a voltage to the $k$th qubit. The 
approach used in this paper, is to apply a permanent
magnetic field with a large gradient in the direction of the spin
chain \cite{g1,g3,g2}. Because of the magnetic field gradient, the
magnetic field is different for different spins, so
that each spin has its unique Larmor frequency.

However, the differences in Larmor frequencies of different spins
do not completely solve the problem of selective excitations. Even
if these differences are large, there are small probabilities of
excitations of unwanted spins with large detunings. This problem
is not very important in conventional nuclear magnetic resonance
spectroscopy nor for a quantum computer with a small number of
qubits, since the pulse sequences do not contain a very large number of
pulses. Unlike this, in a quantum computer with sufficiently large
number of qubits the implementation of logic operations requires
thousands of pulses. The errors caused by the nonresonant
transitions, being small for a single pulse, in general grow
linearly~\cite{Method} with the number of pulses, so that
eventually accumulation of this type of error would impose severe
restrictions on the number of qubits in a nuclear or electron spin
quantum computer. This problem is especially acute because there 
are no error correction codes for correction of this type of
error. 

In this paper we develop an approach which allows one to decrease
significantly (by 50\%) the probabilities of unwanted nonresonant
transitions in order to decrease the errors caused by the {\it rf}
pulses of quantum protocols. Our approach to this
problem is similar to the method used for
minimization of errors caused by the near-resonant
transitions~\cite{Lopez}. Since the probabilities of unwanted
nonresonant transitions oscillate in time with frequencies
proportional to the differences in the Larmor frequencies between
the spins, one can select the optimal values of these frequency
differences (by applying a specific magnetic field
gradient) to minimize the nonresonant transitions. In spite of a
rather refined character of our model (in the sense that this
system is complicated for the experimental realization with a
solid-state device), we believe that our approach provides an
important tool for the solution of similar problems in many
nuclear and electron spin quantum computer proposals.

\section{The Ising spin quantum computer}

The Hamiltonian for an Ising spin chain placed
in an external magnetic field can be written in the form,
\begin{equation}
\label{H}
H_n=-\sum_{k=0}^{L-1}\omega_kI_k^z
-2J\sum_{k=0}^{L-2}I_k^zI_{k+1}^z
-{\Omega_n\over 2}\sum_{k=0}^{L-1}
\left\{I_k^-\exp\left[-i\left(\nu_n t+\varphi_n\right)\right]+
h.c.\right\}
=H_0+V_n(t).
\end{equation}
Here $\hbar=1$; $I_k^\pm=I_k^x\pm I_k^y$;
$I_k^x$, $I_k^y$, and $I_k^z$ are the components of the
operator of the $k$th spin $1/2$; $\omega_k$ is the
Larmor frequency of the $k$th spin; $J$ is the Ising interaction
constant; $\Omega_n$ is the Rabi frequency (frequency of
precession around the resonant transverse field in the rotating
frame); $\nu_n$ and $\varphi_n$ are, respectively, 
the frequency and the phase of the $n$th pulse. 
The Hamiltonian (\ref{H}) is
written for the $n$th rectangular {\it rf} pulse. We assume that
the Larmor frequency difference
$\delta\omega=\omega_{k+1}-\omega_{k}$ between neighboring
spins is independent of the spin's number $k$. Below we omit the
index $n$ which indicates the pulse number. The long-range
dipole-dipole interaction is suppressed by choosing the angle
between the chain and the external permanent magnetic field to be
equal to the magic angle~\cite{Lopez}.

In the interaction representation, the solution of the
Schr\"odinger equation can be written in the form
\begin{equation}
\label{Psi}
\Psi(t)=\sum_nC_q(t)|q\rangle\exp(-iE_qt),
\end{equation}
where $E_q$ and $|q\rangle$ are, respectively,
the eigenvalues and eigenfunctions of the Hamiltonian $H_0$.
The expansion coefficients satisfy the system of linear
differential equations,
\begin{equation}
\label{differential}
i\dot C_q(t)=-{\Omega\over 2}\sum_lC_l(t)|l\rangle
\exp[i(E_q-E_l-\sigma\nu)t-i\sigma\varphi],
\end{equation}
where $\sigma=1$ if $E_q>E_l$ and $\sigma=-1$ if
$E_q<E_l$. The states $|l\rangle$ and $|q\rangle$
in Eq. (\ref{differential}) are related by a flip of
one $k'$th spin, so that the total number of terms in the
right-hand side of Eq. (\ref{differential}) is equal to
the number of qubits, $L$.

If the condition
\begin{equation}
\label{inequality}
\Omega<J\ll\delta\omega
\end{equation}
is satisfied, the pulse effectively affects
only one $k$th spin in the chain~\cite{perturbation}
whose frequency $\omega_k$ is
close (near-resonant)
or equal (resonant) to the frequency of the pulse, $\nu$.
In this approximation, one has to keep only one term
(of the $L$ terms)
on the right-hand side of Eq.~(\ref{differential}) and
the system of coupled differential
equations (\ref{differential}) splits into $2^{L}/2$
independent pairs of equations~\cite{perturbation,1000} of the form
\begin{equation}
\label{dif_dynamics}
i\dot C_m(t)=-{\Omega\over 2}
e^{-i\left(\Delta^{pm} t-\varphi\right)}C_p(t),
\end{equation}
$$
i\dot C_p=-{\Omega\over 2}
e^{i\left(\Delta^{pm} t-\varphi\right)}C_m(t).
$$
Here the states $|m\rangle$ and $|p\rangle$ are related by
a flip of the $k$th spin,
$\Delta^{pm}=E_p-E_m-\nu$, and we suppose that $E_p>E_m$.

\section{Nonresonant transitions}
In this Section, we consider the quantum dynamics which causes
the nonresonant effects. To simplify the notations, we fix the
states $|p\rangle$ and $|m\rangle$, the number of the resonant
spin $k$, and define
\begin{equation}
\label{delta} \delta=E_p-E_m-\nu=\Delta^{pm}.
\end{equation}
The solution of Eq. (\ref{dif_dynamics}) is
\begin{equation}
\label{2x2}
C_m(t_0+\tau)=\left[\cos\left({\lambda\tau\over 2}\right)+
i{\delta\over\lambda}\sin\left({\lambda\tau\over 2}\right)\right]
e^{-i{\delta\over 2}\tau},
\end{equation}
$$
C_p(t_0+\tau)=i{\Omega\over\lambda}
\sin\left({\lambda\tau\over 2}\right)
e^{i\delta t_0-i\varphi+i{\delta\over 2}\tau}.
$$
Here $t_0$ is the time of the beginning of the pulse;
$\lambda=\sqrt{\delta^2+\Omega^2}$; $\tau=t-t_0$ is the
duration of the pulse; and the initial conditions are
\begin{equation}
\label{initial100}
C_m(t_0)=1,\qquad C_p(t_0)=0.
\end{equation}

The solution for the transition from the upper state to the lower
state is
$$
C_m(t_0+\tau)=i{\Omega\over\lambda}
\sin\left({\lambda\tau\over 2}\right)
e^{-i\delta t_0+i\varphi-i{\delta\over 2}\tau},
$$
\begin{equation}
\label{2x2a}
C_p(t_0+\tau)=\left[\cos\left(\lambda\tau\over 2\right)-
i{\delta\over\lambda}
\sin\left({\lambda\tau\over 2}\right)\right]
e^{i{\delta\over 2}\tau},
\end{equation}
$$
C_m(t_0)=0,\qquad C_p(t_0)=1.
$$

Suppose that initially only one state is populated, for example,
\begin{equation}
\label{states}
C_m(t_0)=1,~~C_p(t_0)=0,~~
|m\rangle=|0_40_3{\bf 0}_21_10_0\rangle,~~
|p\rangle=|0_40_3{\bf 1}_21_10_0\rangle~~(k=2).
\end{equation}
The dynamics of the coefficients $C_m(t)$ and $C_p(t)$ is defined
by Eq. (\ref{2x2}). Now we calculate the probabilities for
transitions associated with the flip of a spin with nonresonant
frequency, and with flips of two spins: one with resonant and one
with nonresonant frequencies. We consider, for example, the
transitions $|m\rangle\rightarrow|i\rangle$ and
$|m\rangle\rightarrow|j\rangle$, where
\begin{equation}
\label{states1}
|i\rangle=|0_41_3{\bf 0}_21_10_0\rangle,~~
|j\rangle=|0_41_3{\bf 1}_21_10_0\rangle.
\end{equation}
Here $k'=3$ is the number of the spin with the nonresonant
frequency, and the initial conditions are
\begin{equation}
\label{initial10}
C_i(t_0)=0,~~C_j(t_0)=0.
\end{equation}
The dynamics of these coefficients is defined by
Eqs.~(\ref{differential}) with two essential terms on the
right-hand side,
$$
i\dot C_i(t)=-\frac\Omega 2e^{-i(\Delta t-\varphi)}C_j(t)
-\frac \Omega 2e^{i[(E_i-E_m-\sigma\nu)t-\sigma\varphi)]}C_m(t),
$$
\begin{equation}
\label{dif_dynamics1}
i\dot C_j(t)=-\frac\Omega 2e^{i(\Delta t-\varphi)}C_i(t)
-\frac \Omega 2e^{i[(E_j-E_p-\sigma\nu)t-\sigma\varphi)]}C_p(t),
\end{equation}
where
\begin{equation}
\label{delta1}
\Delta\equiv E_j-E_i-\nu;
\end{equation}
$E_j>E_i$; $\sigma=1$ if $E_i>E_m$ (and $E_j>E_n$) and $\sigma=-1$
if $E_i<E_m$ (and $E_j<E_n$). For example, for the states in
Eqs.~(\ref{states}) and (\ref{states1}) one has $E_i>E_m$ and
$\sigma=1$. We characterize the nonresonant transition with
(large) detuning $D$ (which in general depends on the kind of
state and spin number $k'$) defined as
\begin{equation}
\label{delta2}
D=E_i-E_m-\sigma\nu-{\delta-\Delta\over 2}.
\end{equation}
Then using Eqs. (\ref{delta}), (\ref{delta1}), and (\ref{delta2})
one can write
\begin{equation}
\label{fDifferences}
E_i-E_m-\sigma\nu=D+{\delta-\Delta\over 2},~~~
E_j-E_p-\sigma\nu=D-{\delta-\Delta\over 2}.
\end{equation}
It is convenient to introduce new coefficients
\begin{equation}
\label{nCoefficients}
C_i(t)=e^{-i\frac\Delta 2t}A_i,~~~
C_j(t)=e^{i\frac\Delta 2t-i\varphi}A_j.
\end{equation}
Then one obtains a system of two coupled
differential equations for the coefficients $A_i(t)$ and $A_j(t)$
with an effective oscillating external force,
$$
i\dot A_i(t)+\frac\Delta 2A_i(t)+\frac\Omega 2A_j(t)=
-\frac \Omega 2e^{i\left[\left(D+\frac\delta 2\right)t-\sigma\varphi
\right]}C_m(t),
$$
\begin{equation}
\label{dif_dynamics2}
i\dot A_j(t)-\frac\Delta 2A_j(t)+\frac\Omega 2A_i(t)=
-\frac \Omega 2e^{i\left[\left(D-\frac\delta 2\right)t-
(\sigma-1)\varphi\right]}C_p(t),
\end{equation}
$$
A_i(t_0)=0,~~~~A_j(t_0)=0,
$$
where $C_m(t)$ and $C_p(t)$ are defined by Eq. (\ref{2x2}).
In order to remove the constant phases
from Eq. (\ref{dif_dynamics2}) we write
\begin{equation}
\label{nCoefficients1}
A_i(t)=
e^{i\left[\left(D+\frac\delta 2\right)t_0-\sigma\varphi\right]}
B_i(\tau),~~~
A_j(t)=
e^{i\left[\left(D+\frac\delta 2\right)t_0-\sigma\varphi\right]}
B_j(\tau).
\end{equation}
The coefficients $B_i(\tau)$ and $B_j(\tau)$ 
satisfy the following
system of two coupled differential equations:
$$
i\dot B_i(\tau)+\frac\Delta 2B_i(\tau)+\frac\Omega 2B_j(\tau)=
f(\tau),
$$
\begin{equation}
\label{dif_dynamics3}
i\dot B_j(\tau)-\frac\Delta 2B_j(\tau)+\frac\Omega 2B_i(\tau)=
g(\tau),
\end{equation}
$$
B_i(0)=0,~~~~B_j(0)=0,
$$
where the differentiation is performed with respect to
time-interval, $\tau$, and
$$
f(\tau)=-\frac\Omega 4\left(1+\frac\lambda \delta\right)
e^{i\left(D+\frac\lambda 2\right)\tau}-
\frac\Omega 4\left(1-\frac\lambda \delta\right)
e^{i\left(D-\frac\lambda 2\right)\tau},
$$
\begin{equation}
\label{fg}
g(\tau)=-\frac{\Omega^2}{4\lambda}
\left[e^{i\left(D+\frac\lambda 2\right)\tau}-
e^{i\left(D-\frac\lambda 2\right)\tau}\right].
\end{equation}
By differentiation of the system of equations 
(\ref{dif_dynamics3}),
one obtains two uncoupled second order differential equations,
$$
\ddot B_i(\tau)+\frac{\Lambda^2} 4B_i(\tau)=
\frac\Delta 2f(\tau)+\frac\Omega 2g(\tau)-i\dot f(\tau),
$$
\begin{equation}
\label{dynamics_final100}
B_i(0)=0,~~~~\dot B_i(0)=i\frac\Omega 2
\end{equation}
and
$$
\ddot B_j(\tau)+\frac{\Lambda^2} 4B_j(\tau)=
-\frac\Delta 2g(\tau)+\frac\Omega 2f(\tau)-i\dot g(\tau),
$$
\begin{equation}
\label{dynamics_final10}
B_j(0)=0,~~~~\dot B_j(0)=0,
\end{equation}
where the initial conditions for $\dot B_i(0)$ and $\dot B_j(0)$
follow from Eq. (\ref{dif_dynamics3}) with $f(0)=-\Omega/2$ and
$g(0)=0$.

The solution for both coefficients $B_i(\tau)$
and $B_j(\tau)$ has the form,
$$
a_1e^{i\frac\Lambda 2\tau}+a_2e^{i\frac\Lambda 2\tau}+
a_3e^{i\left(D+\frac\lambda 2\right)\tau}+
a_4e^{i\left(D-\frac\lambda 2\right)\tau}
$$
with different coefficients $a_1,\dots, a_4$.
Keeping only first order terms in $\Omega/D$,
one obtains,
$$
B_i(\tau)=-{\Omega\over 2D}
\left\{\cos\left({\Lambda \tau\over 2}\right)-
\left[\cos\left({\lambda \tau\over 2}\right)+
i\frac\delta\lambda\sin\left({\lambda \tau\over 2}\right)\right]
e^{iD\tau}\right\},
$$
\begin{equation}
\label{dynamics}
B_j(\tau)=-i{\Omega\over 2D}
\left\{\frac\Omega\Lambda\sin\left({\Lambda \tau\over 2}\right)-
\frac\Omega\lambda\sin\left({\lambda \tau\over 2}\right)
e^{iD\tau}\right\}.
\end{equation}
This solution is written for the initial condition
(\ref{initial100}). For the initial conditions given by the third
equation (\ref{2x2a}), the solution is
$$
B_i^\prime(\tau)=-i{\Omega\over 2D}
\left\{\frac\Omega\Lambda\sin\left({\Lambda \tau\over 2}\right)-
\frac\Omega\lambda\sin\left({\lambda \tau\over 2}\right)
e^{iD\tau}\right\},
$$
\begin{equation}
\label{dynamics1}
B_j^\prime(\tau)=-{\Omega\over 2D}
\left\{\cos\left({\Lambda \tau\over 2}\right)-
\left[\cos\left({\lambda \tau\over 2}\right)-
i\frac\delta\lambda\sin\left({\lambda \tau\over 2}\right)\right]
e^{iD\tau}\right\}.
\end{equation}
Here the coefficients $B_i^\prime(\tau)$ and $B_j^\prime(\tau)$
are obtained using Eq. (\ref{2x2a}) and the properly modified
relations (\ref{nCoefficients}) and (\ref{nCoefficients1}) (which
result in different common phase factors, unimportant for further
consideration).

\section{Probabilities of nonresonant transitions}

The nonresonant transitions considered in the previous Section
generate errors in the quantum computer. Our objective is to
minimize the probabilities of these transitions (for estimations
of these probabilities see
Refs.~\cite{perturbation,1000,Felix1}) as much
as possible by choosing optimal parameters of the model. As follows
from Eq. (\ref{dynamics}), the probability of a nonresonant
transitions associated with the flip of a spin 
with nonresonant
frequency is proportional to $(\Omega/2\delta\omega)^2$. 
(See also
Refs. \cite{perturbation,1000}.) If one could suppress these
transitions, the total probability of error would be associated
with flips of two spins with nonresonant frequencies, and it
would be of order of $(\Omega/2\delta\omega)^4$. Successful
solution of this problem would allow one 
to substantially decrease
the errors introduced by the {\it rf} pulses and to relieve the
requirements for large magnetic field gradients characterized
by $\delta\omega$.

In order to estimate the errors given by Eq. (\ref{dynamics}),
we will analyze this equation
for a typical example. Consider the errors generated during
implementation of the (useful) transition
\begin{equation}
\label{useful}
|m\rangle\rightarrow|p\rangle,
\end{equation}
where the states
$|m\rangle$ and $|p\rangle$ are defined in Eq. (\ref{states}).
Since the frequency of the external {\it rf} field is resonant
for this transition, then
\begin{equation}
\label{resonantP}
\nu=E_p-E_m=3\delta\omega,~~~\delta=0~~~\lambda=\Omega,~~~
\tau={\pi\over\Omega}~~~k=2,
\end{equation}
where $k$ is the number of the spin with the resonant
frequency.

The values
\begin{equation}
\label{Delta_Lambda}
\Delta_{k'},~~~~\lambda_{k'}=\sqrt{\Delta^2_{k'}+\Omega^2}
\end{equation}
for nonresonant transitions to the states with one
flipped $k'$ spin are
\begin{equation}
\label{Deltas}
\Delta_{k'}=\delta,~~~\Lambda_{k'}=\lambda~~~{\rm if}~~~k'\ne k\pm 1.
\end{equation}
When $k'=\pm 1$ the detunings are different,
$\Delta_{k'}\ne\delta$, and depend on the orientations of the
$(k+1)$th and $(k-1)$th spins in the state $|i\rangle$. For
example, for the states in Eq. (\ref{states1}) one has
$E_j-E_i=3\delta\omega-2J$ so that
\begin{equation}
\label{Deltas1}
\Delta_3=E_j-E_i-\nu=\delta-2J=-2J.
\end{equation}
For the detunings associated with the other spins, one has
\begin{equation}
\label{Deltas2}
\Delta_0=0,~~~\Delta_1=2J~~~\Delta_4=0.
\end{equation}

In order to produce resonant transitions (where $\delta=0$),
the time-interval should be 
$\tau=\pi/\Omega$ [$\pi$-pulse, see Eq. (\ref{2x2})]. 
In order to suppress possible
near-resonant transitions, the value of $\Omega$ should satisfy
the $2\pi K$-condition~\cite{Lopez}
\begin{equation}
\label{2pik}
\Omega^{(K)}={2J\over\sqrt{4K^2-1}}.
\end{equation}
For these parameters and for the state $|m\rangle$ in
Eq. (\ref{states}) (for $k=2$) the quantities in
Eq. (\ref{dynamics}) take the following values:
$$
\delta=0,~~~\lambda=\Omega,~~~
\sin\left({\lambda\tau\over 2}\right)=1,~~~
\cos\left({\lambda\tau\over 2}\right)=0,
$$
\begin{equation}
\label{quantities}
\sin\left({\Lambda_{k'}\tau\over 2}\right)=1,~~~
\cos\left({\Lambda_{k'}\tau\over 2}\right)=0,~~~{\rm for}~~~
{k'\ne k\pm 1},
\end{equation}
$$
\sin\left({\Lambda_{k'}\tau\over 2}\right)=0,~~~
\cos\left({\Lambda_{k'}\tau\over 2}\right)=1,~~~{\rm for}~~~
{k'=k\pm 1}.
$$
Then, after the $\pi$-pulse the coefficients in
Eq. (\ref{dynamics}) become
\begin{equation}
\label{dynamics10}
B_i^{k'}=0,~~~B_j^{k'}=-i{\Omega\over 2D_{k'}}
\left(1-e^{i\pi {D_{k'}\over\Omega}}\right)~~~
{\rm for}~~~ k'\ne k\pm 1,
\end{equation}
and
\begin{equation}
\label{dynamics11}
B_i^{k'}=-{\Omega\over 2D_{k'}},~~~
B_j^{k'}=
i{\Omega\over 2D_{k'}}e^{i\pi {D_{k'}\over\Omega}}~~~
{\rm for}~~~ k'=k\pm 1,
\end{equation}
where the upper index of the amplitudes $B_i^{k'}$ and $B_j^{k'}$
indicates that these states are associated with the initial state 
by the flip of the $k'$th spin.

The coefficients $B_i^{k'}$ and $B_j^{k'}$
in Eqs. (\ref{dynamics10}) and
(\ref{dynamics11}) are calculated for a transition resonant
for the initial state.
If the frequency of the {\it rf} pulse is close to resonant
(near-resonant), then
$$
\delta\ne 0,~~~\lambda=\sqrt{\Omega^2+4J^2},~~~
\sin\left({\lambda\tau\over 2}\right)=0,~~~
\cos\left({\lambda\tau\over 2}\right)=1,
$$
\begin{equation}
\label{quantities1}
\sin\left({\Lambda_{k'}\tau\over 2}\right)=0,~~~
\cos\left({\Lambda_{k'}\tau\over 2}\right)=1,~~~{\rm for}~~~
{k'\ne k\pm 1},
\end{equation}
$$
\sin\left({\Lambda_{k'}\tau\over 2}\right)=1,~~~
\cos\left({\Lambda_{k'}\tau\over 2}\right)=0,~~~{\rm for}~~~
{k'=k\pm 1}.
$$

\begin{figure}
\includegraphics[width=11cm,height=8cm]{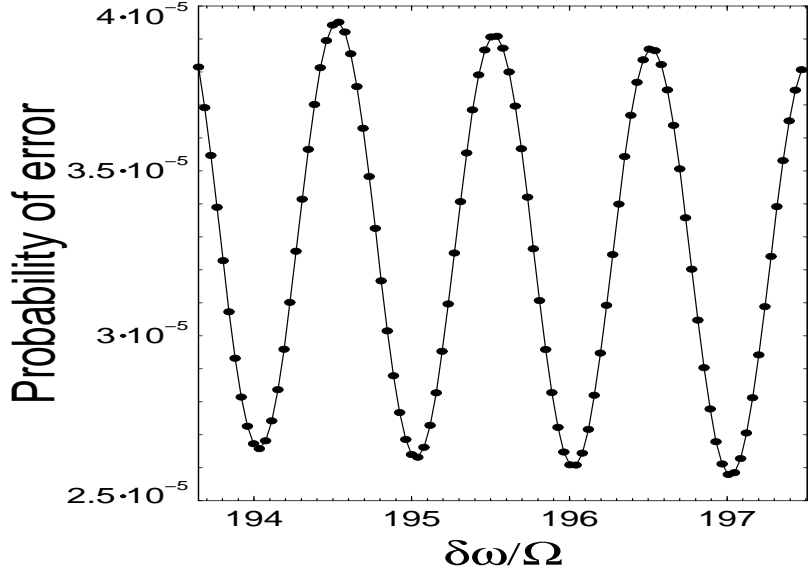}
\vspace{-5mm}
\caption{Probability of errors after application of $\pi$-pulse
resonant for the transition (\ref{useful}) for states
and initial conditions defined in Eq. (\ref{states}). The solid
line indicates the analytic results calculated using
Eq. (\ref{Pk}).The filled circles are calculated using exact numerical
solution~\cite{perturbation,1000}. $L=5$,
$\Omega/J=\Omega^{(K)}/J\approx 1/2$, $K=2$ [see Eq. (\ref{2pik})].}
\label{fig:1}
\end{figure}

The error $P_k$ for an arbitrary initial state generated by the
pulse resonant for the transition (\ref{transition}) is the sum of
the probabilities of all nonresonant transitions associated with
flips of all spins with $k'\ne k$,
\begin{equation}
\label{Pk}
P_k=2\left({\Omega\over 2D_{k-1}}\right)^2+
2\left({\Omega\over 2D_{k+1}}\right)^2+
\sum_{k'\ne k-1,k,k+1}
\left({\Omega\over 2D_{k'}}\right)^2
\left|1-e^{i\pi {D_{k'}\over\Omega}}\right|^2.
\end{equation}
If $k=1$ then the term with $D_{k-1}$ in Eq. (\ref{Pk})
should be omitted. If
$k=L-1$, one must omit the term with $D_{k+1}$.
In Fig.~1 we plot the total probability of error after
application of a $\pi$-pulse resonant for the transition
(\ref{useful}) with the states and initial conditions
defined in Eq. (\ref{states}). One can see a good 
correspondence between the numerical and analytical 
results. The same agreement was observed for other 
initial states and parameters of the model.
The parameters used in the simulations can be realized
using, for example, phosphorus impurity donors in silicon.
(For the relation between the numerical and physical parameters,
see, for example, Ref. \cite{Method}.)

\section{Minimization of nonresonant transitions}

From Eq. (\ref{Pk}) and Fig. 1 one can see that probability of
error oscillates as a function of the ratio $\delta\omega/\Omega$.
As follows from Eq. (\ref{Pk}) the period of oscillations is equal
to 2. (In Fig. 1 the period is twice smaller because of the
symmetry of the transition: the probabilities oscillate only 
for the transitions with detunings of the order 
of $2\delta\omega$.)

In order to describe these oscillations
we continue to analyze the example from the previous Section. 
The values of $D_{k'}$ for the initial state $|m\rangle$ in Eq.
(\ref{states}) and for $k=2$ are
\begin{equation}
\label{Dk}
D_0=-2\delta\omega-J,~~D_1=\delta\omega-J,~~D_4=\delta\omega+J,~~
D_5=2\delta\omega+J.
\end{equation}
Under the condition
\begin{equation}
\label{min5}
|{D_0\over\Omega}|=|{D_5\over\Omega}|=2q,
\end{equation}
all (two) terms in the sum in Eq. (\ref{Pk}) vanish, and the
probability of error takes its minimum value equal to
\begin{equation}
\label{minn} P^{\rm
min}\approx\left({\Omega\over\delta\omega}\right)^2 \approx
{1\over q^2}.
\end{equation}
For example, in Fig. 2 for $\delta\omega/\Omega\approx q=194$
one has $P^{\rm min}\approx 2.65\times 10^{-5}$.

\begin{figure}
\includegraphics[width=11cm,height=8cm]{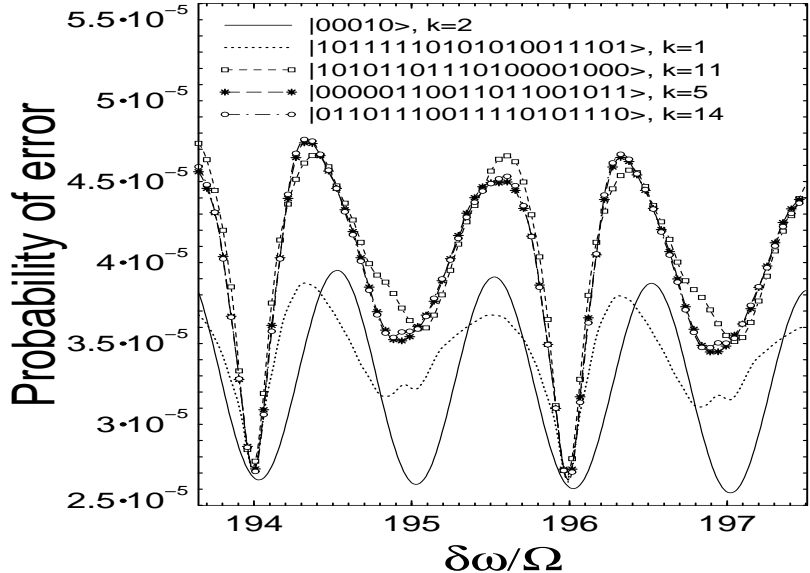}
\vspace{-5mm}
\caption{The same as Fig. 1 but for different states.
The results are obtained using Eq. (\ref{Pk}).}
\label{fig:2}
\end{figure}

We analyzed the nonresonant transitions only for one state. We
now discuss the general case. In Fig. 2 we plot the probabilities
of errors for four randomly selected states generated by the
pulse, resonant for the transition (\ref{transition}) for $L=20$
qubits. For comparison, we also plot the curve from Fig. 1 (solid
line) analyzed before. In principle, one could calculate errors
for an arbitrary number of qubits. However, since the
probabilities of flips of distant qubits decrease as $1/(k-k')^2$,
the distant qubits  do not contribute significantly to the total
error.

From Fig. 2 one can see that errors for different states take
their minimum values at the same values of $\delta\omega/\Omega$.
It is convenient to write the detunings $D_{kk'}$ (here $k$ is the
number of spin with the resonant or near-resonant frequency, and
$k'$ is the number of spin with nonresonant frequency) in the
form
\begin{equation}
\label{minQ1}
|D_{kk'}|=|k-k'|\delta\omega+\alpha_{kk'} J,
\end{equation}
where $\alpha_{kk'}$ can assume the integer 
values $-2,-1,0,1,2$.
Choosing the parameters
\begin{equation}
\label{main_condition}
{\delta\omega\over\Omega}=2Q,~~~K=2{\cal K},
\end{equation}
where $Q\gg 1$ and ${\cal K}$ are integers,
one can write
\begin{equation}
\label{DkOmega}
\left|{D_{kk'}\over\Omega}\right|\approx
2Q|k-k'|+\alpha_{kk'}\left(2{\cal K}-{1\over 16{\cal K}}\right).
\end{equation}
Here we used Eq. (\ref{2pik}).
The value of $Q\gg 1$ defines the magnetic field gradient
characterized by the ratio $\delta\omega/\Omega$,
and ${\cal K}\ge 1$ determines the Rabi frequency, or $\Omega/J$
at a given value of the Ising interaction constant $J$.

\begin{figure}
\includegraphics[width=11cm,height=8cm]{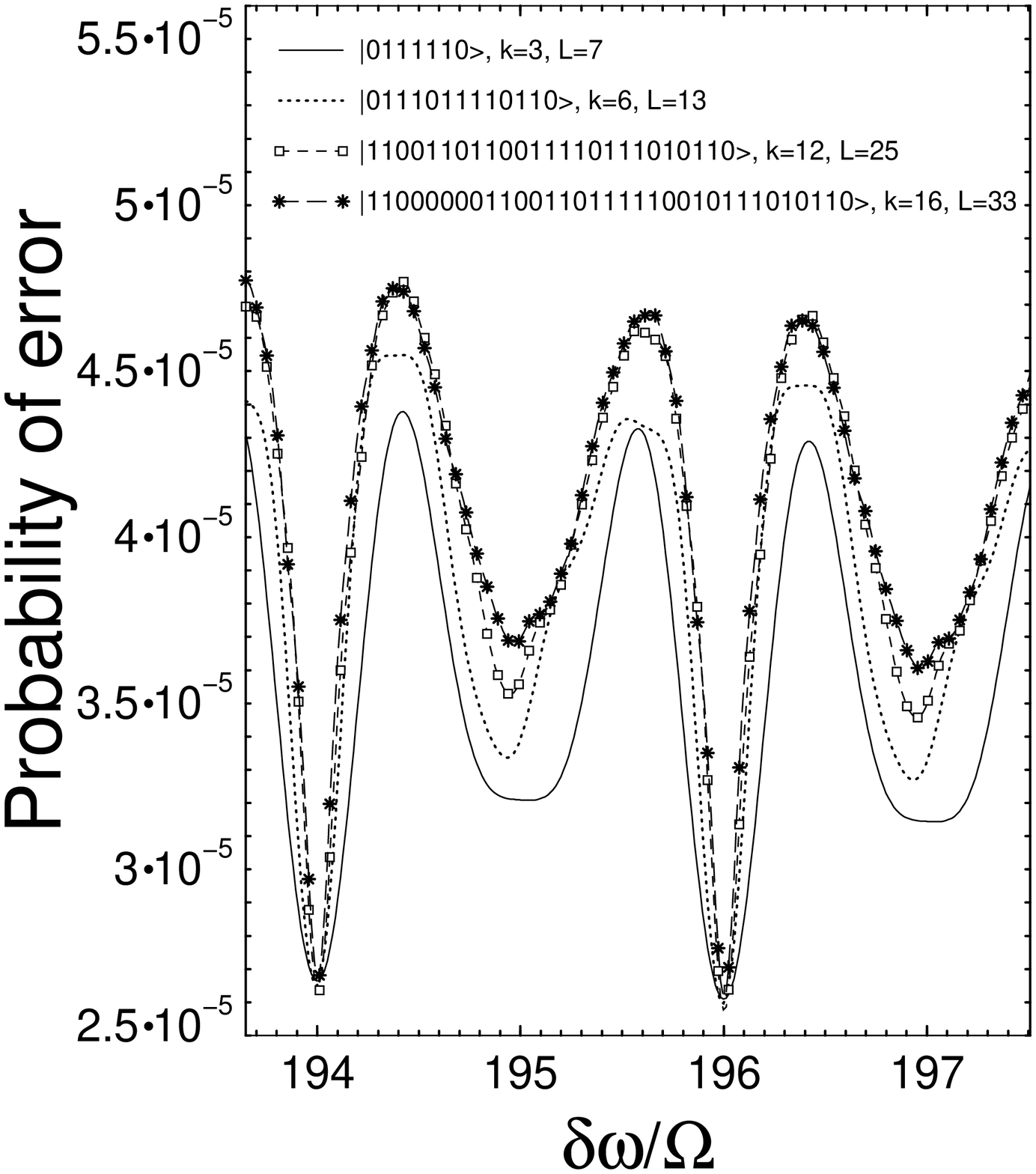}
\vspace{-5mm}
\caption{The error for different number $L$ of spins in
the spin chain. k is chosen in the center of the chain when
the error takes its maximum value. ${\cal K}=1$.}
\label{fig:3}
\end{figure}

Under the conditions (\ref{main_condition})
the errors $P_{kk'}$ associated with flips of $k'$th spins
with $k'\ne k,k\pm 1$ (below called the distant spins)
in the sum of Eq. (\ref{Pk}) take their minimum values equal to
\begin{equation}
\label{PkSmall}
P_{kk'}\approx {1\over 4Q^2}\left({\pi\over 32 {\cal K}}\right)^2
\left({\alpha_{kk'}\over k-k'}\right)^2.
\end{equation}
At the minimums, the transitions associated with the flips of the
distant spins are mostly suppressed, and the total error
(\ref{minn}) is mainly associated with flips of the $(k-1)$th and
the $(k+1)$th spins,
\begin{equation}
\label{Ptotal}
P_{k}^{\rm min}\approx {1\over 4Q^2}\left[
1-{2{\cal K}\over Q}(\alpha_{k,k-1}+\alpha_{k,k+1})
\left(1-{1\over 32{\cal K}^2}\right)+
\left({\pi\over 32 {\cal K}}\right)^2\sum_{k'\ne k-1,k,k+1}
\left({\alpha_{kk'}\over k-k'}\right)^2\right].
\end{equation}
From this equation one can see that even for minimum value
${\cal K}=1$, the error associated with flips of distant
spins is at least two orders
smaller than the error associated
with flips of $(k+1)$th and $(k-1)$th spins, and this
error quickly decreases as ${\cal K}$ increases.

Since the transitions associated with flips of the distant spins
are mostly suppressed, for the optimal parameters
(\ref{main_condition}) the error is practically independent of the
number of qubits. In Fig. 3 we compare the errors generated by
nonresonant transitions for different numbers of spins in the
spin chain. From Fig. 3 one can see that, in general, the
error grows as the number of qubits increases, but the error in
the minima is almost independent of the number of qubits $L$.

\begin{figure}
\mbox{
\includegraphics[width=8cm,height=8cm]{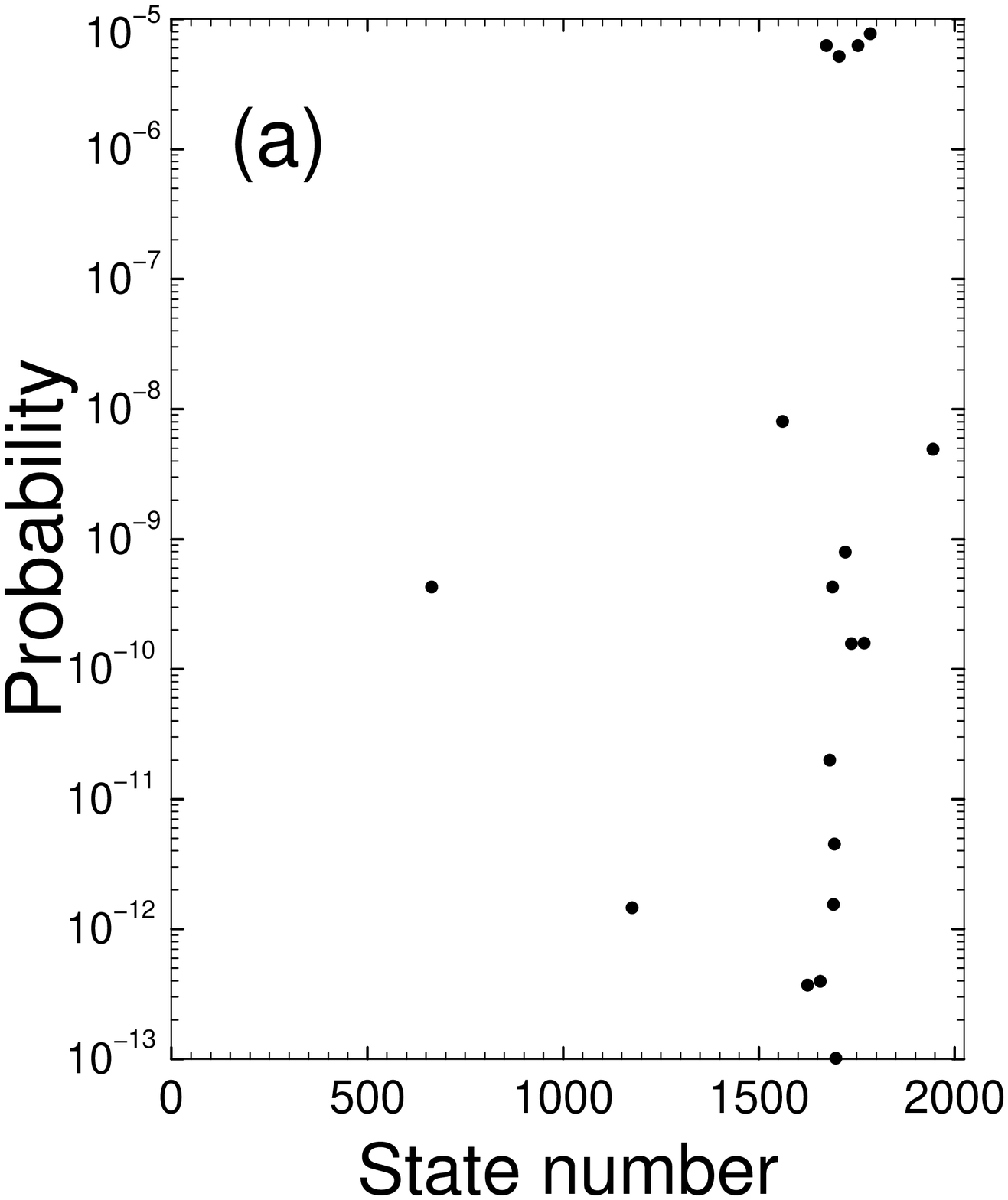}
\includegraphics[width=8cm,height=8cm]{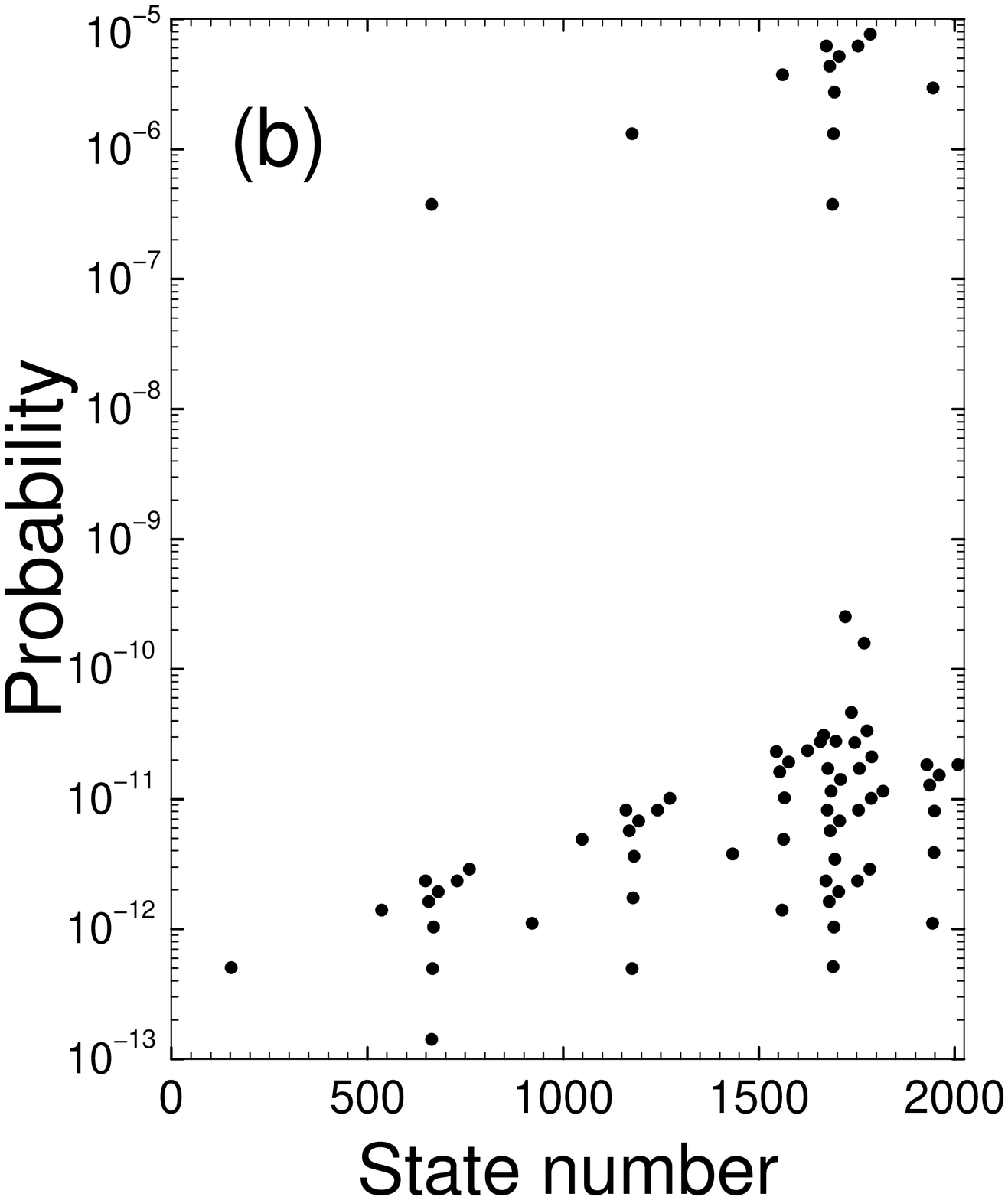}}
\vspace{-5mm} \caption{Probabilities of unwanted states created in
result of the $\pi$-pulse resonant to the transition
(\ref{transition}). Initial state is $|11010111001\rangle$, $k=5$,
$L=11$, ${\cal K}=5$ ($\Omega/J\approx 0.1$); (a)
$\delta\omega/\Omega=200$ ($Q=100$), (b)
$\delta\omega/\Omega=200.314$. The results are obtained using
exact numerical solution \cite{perturbation,1000}.} \label{fig:4}
\end{figure}

It is important that the error is associated only with the
$(k+1)$th, $k$th, and the $(k-1)$th spins. 
The probabilities of flips of other
spins are small when the condition (\ref{main_condition}) is satisfied. In
Figs. 4(a,b) the probabilities of error are plotted for 
different states for the optimal value 
$\delta\omega/\Omega=2Q=200$ [Fig. 4(a)],
and for another value $\delta\omega/\Omega=200.314$ [Fig. 4(b)].
By comparison of Fig. 4(a) and Fig. 4(b), one can see that
optimization of the parameters results in the 
relative suppression of
all but four transitions, associated with flips of $(k-1)$th,
$k$th, and $(k+1)$th spins. Since the errors are associated with the 
definite three spins these errors,
probably, can be corrected by application of error-correction
codes associated with these particular spins.

\section{Conclusion}

In this paper, we considered the effect of suppression of
nonresonant transitions only for one pulse. Since a quantum
protocol consists of many such pulses, application of our results
would substantially decrease {\it the rate} with which the error
accumulates in the register of a quantum computer.

In our model the Larmor frequency $w_0+k'\delta\omega$
of the $k'$th spin increases with increasing spin number $k'$.
If $k$ is the number of the spin with resonant frequency,
then the moduli of detunings $\delta\omega|k-k'|$ increase
with increasing ``distance'' $|k-k'|$, so that the
error associated with flips of distant spins
decreases as $1/(k-k')^2$.
The situation is different for Kane's quantum computer
\cite{Kane}. In this computer the Larmor frequencies are the same
for different spins, except for one spin whose frequency is
changed by application of a voltage to this particular spin.
Hence, the detunings
are the same for all spins in the
chain (except for the one spin to which the voltage is applied)
and equal to $\delta\omega$.
As a result, the error caused by the
nonresonant transitions should increase linearly with an increase in
the number of qubits. Our method would allow one to 
suppress this linear growth.

\begin{acknowledgments}
We thank to G. D. Doolen for useful discussions.
This work was supported by the Department of Energy (DOE) under
Contract No. W-7405-ENG-36, by the National Security Agency (NSA),
and by the Advanced Research and Development Activity (ARDA).
\end{acknowledgments}

{}
\end{document}